\documentclass[useAMS,usenatbib]{mn2e}
\usepackage{graphicx}
\usepackage{aas_macros}
\usepackage{amssymb}
\usepackage{url}

\newcommand{\Ld}{\ensuremath{L_{\mathrm{D}}}}
\newcommand{\Lpl}{\ensuremath{L_{\mathrm{PL}}}}
\newcommand{\Lrad}{\ensuremath{L_{\mathrm{R}}}}
\newcommand{\Lx}{\ensuremath{L_{\mathrm{X}}}}
\newcommand{\Lopt}{\ensuremath{L_{\mathrm{O}}}}
\newcommand{\Rlo}{\ensuremath{R_{\mathrm{low}}}}
\newcommand{\Rhi}{\ensuremath{R_{\mathrm{high}}}}

\title[Accretion states and radio loudness in AGN]{Accretion states
  and radio loudness in Active Galactic Nuclei: analogies with X-ray
  binaries} 
\author[Elmar G.\ K\"ording, Sebastian Jester and Rob Fender]{Elmar
  G.\ K\"ording$\thanks{E-mail: Elmar@phys.soton.ac.uk}$, Sebastian
  Jester$\thanks{Otto Hahn Fellow}$, Rob Fender\\ School of Physics
  and Astronomy, University of Southampton, Southampton SO17 1BJ,
  United Kingdom }
\begin{document}

\date{Accepted 2006 August 17 Received 2006 August 17 in original form 2006 June 23}

\pagerange{\pageref{firstpage}--\pageref{lastpage}} \pubyear{2006}

\maketitle

\label{firstpage}

\begin{abstract}
Hardness-intensity diagrams (HIDs) have been used with great success
to study the accretion states and their connection to radio jets in
X-ray binaries (XRBs).  The analogy between XRBs and active galactic
nuclei (AGN) suggests that similar diagrams may help to understand and
identify accretion states in AGN and their connection to
radio loudness. We construct ``disc-fraction luminosity diagrams''
(DFLDs) as a generalization of HIDs, which plot the intensity against
the fraction of the disc contribution in the overall spectral energy
distribution (SED).  Using a sample of 4963 Sloan Digital Sky Survey
(SDSS) quasars with ROSAT matches, we show empirically that an AGN is
more likely to have a high radio:optical flux ratio when it has a high
total luminosity or a large non-thermal contribution to the SED.  We
find that one has to consider at least two-dimensional diagrams to
understand the radio loudness of AGN.  To extend our DFLD to lower
luminosities we also include a sample of low-luminosity AGN.  Using a
simulated population of XRBs we show that stellar and supermassive BHs
populate similar regions in the DFLD and show similar radio/jet
properties.  This supports the idea the AGN and XRBs have the same
accretion states and associated jet properties.
\end{abstract}


\section{Introduction}
In spite of a vast difference of scales, both black-hole X-ray
binaries (XRBs) and active galactic nuclei (AGN) are thought to be
powered by an essentially scale-invariant central engine consisting of
a black hole, an accretion disc surrounded by a corona and a
relativistic jet
\citep[e.g.,][]{ShakuraSunyaev1973,Antonucci1993,MirabelRodriguez1999}.
In the study of AGN, it is a long-standing puzzle that some objects
show strong radio emission while in others only little radio emission
is found, even though their optical properties are similar
\citep{KellermannSramekSchmidt1989,MillerRawlingsSaunders1993}.  XRBs,
i.e., stellar-mass accreting black holes, show states associated with
strong radio emission as well as states during which the radio
emission is strongly suppressed or ``quenched''
\citep{TananbaumGurskyKellogg1972,FenderCorbelTzioumis1999,CorbelFenderTzioumis2000,Fender2001}.
The different levels of radio emission have been ascribed to different
jet properties of the different accretion states
\citep[e.g.,][]{FenderBelloniGallo2004}.  Due to the shorter
timescales of accretion onto XRBs, it is possible to study the
different accretion states and their association with jets in detail.

The quantitative comparison of XRBs and AGN has made a major step
forwards in recent years with the discovery of a \emph{fundamental
plane of black-hole activity}
\citep{MerloniHeinzdiMatteo2003,FalckeKoerdingMarkoff2004}. This is a
correlation between radio and X-ray luminosities and black hole mass
which fits both classes of objects (both individually and as an
ensemble).  Also the variability properties of AGN and XRBs show many
similarities
\citep{UttleyMcHardyPapadakis2002,MarkowitzEdelsonVaughan2003,KoerdingFalcke2004,AbramowiczKluzniakMcClintock2004}.
Therefore, it is a promising route to use the concepts and methods
developed to study the evolution of radio jets in XRBs for exploring
the puzzle of the radio loudness in AGN.

XRB outbursts seem to occur in cycles of several distinct states that
are similar in every large outburst and from object to object (e.g.,
\citealt{Nowak1995,BelloniHomanCasella2005},see Fig.~\ref{fiHIDSketch}).  At the beginning of the
cycle the source is in the \emph{hard state} characterised by a hard
power law in the X-ray spectrum. At radio frequencies, one ususally
observes a steady jet \citep{Fender2001}. In this state the standard
accretion disc \citep{ShakuraSunyaev1973} is probably truncated 
or unobservable for some other reason and
the inner part of the accretion disc replaced by an inefficient
accretion flow (e.g., an advection dominated accretion flows, ADAF:
\citealt{NarayanYi1994,ChenAbramowiczLasota1995,EsinMcClintockNarayan1997}
but see also \citealt{MillerHomanSteeghs2006}). The X-rays are ususally
described with Comptonization models
(e.g. \citealt{ThornePrice1975,SunyaevTruemper1979}). However, it is
also possible that the compact jet may contribute to the X-ray
emission or even dominate it
\citep{MarkoffFalckeFender2001,MarkoffNowakWilms2005}. Once the source
leaves the hard state, it is often found in an \emph{intermediate
state} (IMS). One finds two IMSs: first, the source enters a
\emph{hard IMS} characterised by a hard spectral component and
band-limited noise in the power spectrum.  Also in this state a jet is
observed in the radio. Later, a \emph{soft IMS} is found, dominated by
a soft spectral component with power-law noise in the power spectrum
(e.g., \citealt{BelloniHomanCasella2005}).  Approximately at the transition
from the soft to the hard IMS one usually finds a radio flare and the
ejecta can often be resolved in the radio band.
In the soft IMS no radio
emission is found. After leaving the IMS, the source may go to the
\emph{soft state}, where the X-ray spectrum is dominated by a soft
thermal component that is thought to originate from a standard
accretion disc. The radio jet is still quenched in this state
\citep{FenderCorbelTzioumis1999,CorbelFenderTzioumis2000}.

It has not been established definitely how the phenomenology of AGN
can be mapped onto these XRB states. Quasars are AGN with bright
optical continuum emission, which are thought to have high accretion
rates and strong disc emission \citep[e.g.,][]{UrryPadovani1995}.
Here, we use the term AGN in a general sense to mean all supermassive
accreting black holes, and we use the term ``quasar'' irrespective of
the radio properties, i.e., synonymous with ``quasi-stellar object''
(QSO), and without imposing any luminosity cut.

For AGN, one commonly adopted definition of a ``radio-loud'' source is
for the $R$ parameter, the ratio of the radio flux (at 1.4 GHz) to the
optical $B$-band flux, to be larger than 10.  We follow this
convention here. It has been known for over two decades that
radio-loud quasars have higher X-ray fluxes that radio-quiet objects
(e.g.,
\citealt{ZamoraniHenryMaccacaro1981,ElvisWilkesMcDowell1994,ZdziarskiJohnsonDone1995}),
perhaps hinting (with hindsight) that they represent different
accretion states.  \citet{PoundsDoneOsborne1995} suggested that
narrow-line Seyfert 1 objects may be analogues of the soft state. The
fundamental plane of accreting black holes has been interpreted as
evidence that low-luminosity AGN (LLAGN) are hard-state objects
\citep{FalckeKoerdingMarkoff2004, KoerdingFalckeCorbel2005}. This has
also been suggested by \citet{Ho2005}, based on the lack of an
observable ``big blue bump'' in the spectral energy distributions
(SEDs) of LLAGN. Again using the fundamental plane,
\citet{MaccaroneGalloFender2003} suggested that AGN with an
intermediate accretion rate (a few percent of the Eddington rate) are
the analogue of soft-state XRBs.

On the other hand, \citet{Boroson2002} has suggested that radio-loud
and radio-quiet quasars occupy distinct regions of a two-dimensional
diagram of physical accretion rate $\dot{M}$ against Eddington-scaled
luminosity $L/L_\mathrm{Edd}$.  \citeauthor{Boroson2002} considers
these two quantities as the likely physical origin of the
``eigenvectors'' 1 and 2 \citep{BorosonGreen1992}, and eigenvector 1
appears to be related to the radio properties.  The division between
radio-loud and radio-quiet objects proposed by \citet{Boroson2002} is
one in black-hole mass, similar to those reported elsewhere
\citep{FranceschiniVercelloneFabian1998,LacyLaurent-MuehleisenRidgway2001,JarvisMcLure2002}.
However, neither \citet{Ho2002} nor \citet{WooUrry2002b} found any
evidence for a mass threshold between radio-loud and radio-quiet
objects.  Thus, the physical mechanism making some sources much more
radio-loud than others remains a puzzle.

For X-ray binaries, hardness-intensity diagrams (HIDs) have been used
with great success to study the evolution of outbursts and separate
the different accretion states (e.g,
\citealt{HomanWijnandsBelloni2001,FenderBelloniGallo2004,BelloniHomanCasella2005}).
The radio-loud and radio-quiet phases of a source are well separated
on this HID. Additionally, the timing properties of the X-ray
lightcurve change dramatically with the position in the HID (e.g.,
\citealt{HomanWijnandsBelloni2001,BelloniHomanCasella2005,Remillard2005}). Hence,
the construction of equivalent diagrams promises to be a useful tool
for the study of AGN, too.  For XRBs, X-ray HIDs are useful diagnostic
tools because spectral features of both the accretion disc and the
hard power-law component are visible in the X-rays. The blackbody
temperature of the accretion disc scales with black hole mass as
$M^{-1/4}$ so that the SEDs of accretion discs around supermassive
black holes peak in the optical/ultra-violet and an X-ray HID for AGN
would not contain any information about the disc emission. Therefore,
we have to construct a new type of diagram which is applicable to the
study of XRBs as well as of AGN. In this paper, we present the
definition of an analogous diagram for AGN (\S\ref{s:method}) and use
our method to explore the origin of radio loudness in a sample of
quasars from the Sloan Digital Sky Survey (SDSS) and LLAGN
(\S\ref{s:results}), also considering the impact of selection effects
and other biases (\S\ref{s:critical}).  We compare the behaviour of
AGN and X-ray binaries in \S\ref{s:disc} and conclude in
\S\ref{s:conc}.

\section{Method}
\label{s:method}

\subsection{Scaling hardness flux diagrams from XRBs to AGN}
\begin{figure}
\resizebox{8.7cm}{!}{\includegraphics{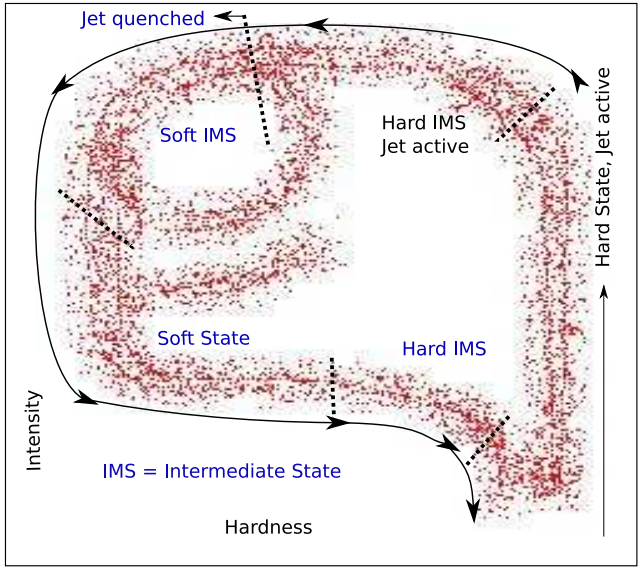}}
\caption{Sketch of an HID found for XRBs. For the abscissa one usually
uses the hardness ratio and the ordinate represents the measured X-ray
counts, i.e., the total measured X-ray flux. The arrows indicate the
movement of a source during outburst.}
\label{fiHIDSketch}
\end{figure}

X-ray HIDs are used to study the evolution of an XRB during an
outburst \citep{FenderBelloniGallo2004,BelloniHomanCasella2005}. On
the ordinate ($y$-axis) of these diagrams, one usually plots the X-ray
count rate, while the abscissa ($x$-axis) denotes the hardness ratio
(e.g., 6.3--10.5 keV count rate/3.8--6.3 keV count rate). For a sketch
of a typical HID see Fig.~\ref{fiHIDSketch}.  While a source is in the
\emph{hard state}, its hard power-law, generally assumed to arise via
Comptonization, dominates the spectrum and the source is on the
right-hand (hard) side of the diagram.  The X-ray emission in the
\emph{soft thermally-dominated state} is believed to be mainly due to
a standard thin accretion disc (multi-temperature black-body). In
this state, the source moves towards the left of the diagram.  In AGN,
the thermal emission of the accretion disc peaks in the optical or
ultra-violet --- far from the observed X-ray bands. Thus, an HID based
solely on the X-ray spectrum cannot probe whether an AGN is in a ``soft''
or ``hard'' state.

An HID that can be used to diagnose AGN as well as XRBs has to be
based on parameters that are independent of the observing band, as at
least the temperature of the accretion disc and therefore the peak of
the emission component scales with black hole mass. The location of an
individual source in the HID is determined by the total luminosity of the
system and the relative strength of the power law component compared
to the disc component of a source. In XRBs, both the disc and the
power-law component are observable in the X-ray band. The \emph{X-ray
luminosity} is (caveat bolometric corrections) equal to the luminosity
$\Ld+\Lpl$, where $\Ld$ denotes the disc luminosity and $\Lpl$ the
luminosity of the power-law component. The \emph{hardness} reflects which
component dominates the spectrum. Thus, $\frac{\Lpl}{\Ld+\Lpl}$ has a
similar behaviour as the hardness ratio. We will refer to this
quantity as the \emph{non-thermal fraction}.  This value approaches zero for
disc-dominated sources and 1 if the power law is the dominating
component.

In AGN, the disc luminosity can be measured in the optical band for
non-obscured sources, while X-ray observations provide information
only about the power-law component. A benefit of the term
$\frac{\Lpl}{\Ld+\Lpl}$ is that it stays finite if one of the
components approaches zero, which happens when a source is in the hard
or the soft state. As it is a ratio of luminosities, it is furthermore
independent of the distance. For the case that the luminosities of
both spectral components scale similarly with BH mass, it is also
independent of the BH mass. We will refer to diagrams plotting
$\Ld+\Lpl$ against $\frac{\Lpl}{\Ld+\Lpl}$ as
\emph{disc-fraction/luminosity diagrams} (DFLDs).

We note in passing that the ratio $\frac{\Lpl}{\Ld+\Lpl}$ is probably connected
  to the properties of the accretion disk wind at least in AGN.  The
  UV luminosity ($\Ld$) is a measure of the radiation pressure while
  the X-ray luminosity ($\Lpl$) influences the ionisation state
  \citep[see, e.g.,][and references therein]{Richards2006}. 

For XRBs we can observe the evolution of an entire outburst of an
individual source and obtain a closed loop on the HID or DFLD, as the
timescales of an outburst is months to years
\citep{ChenShraderLivio1997}. For each position in the diagram we can
observe whether or not the jet is visible. However, for AGN it is
impossible to observe a full outburst cycle (quiescent $\rightarrow$ quasar phase $\rightarrow$ quiescent) as it would last for
hundreds of millions of years (assuming a linear scaling of timescales
with BH mass).  Instead, we have to study a large population of AGN in
the DFLD and observe their average radio loudness (i.e., the
radio:optical flux ratio) as a tracer of the jet activity.

\subsection{Our sample}
\label{s:method.sample}
\begin{figure*}
\resizebox{8.7cm}{!}{\includegraphics{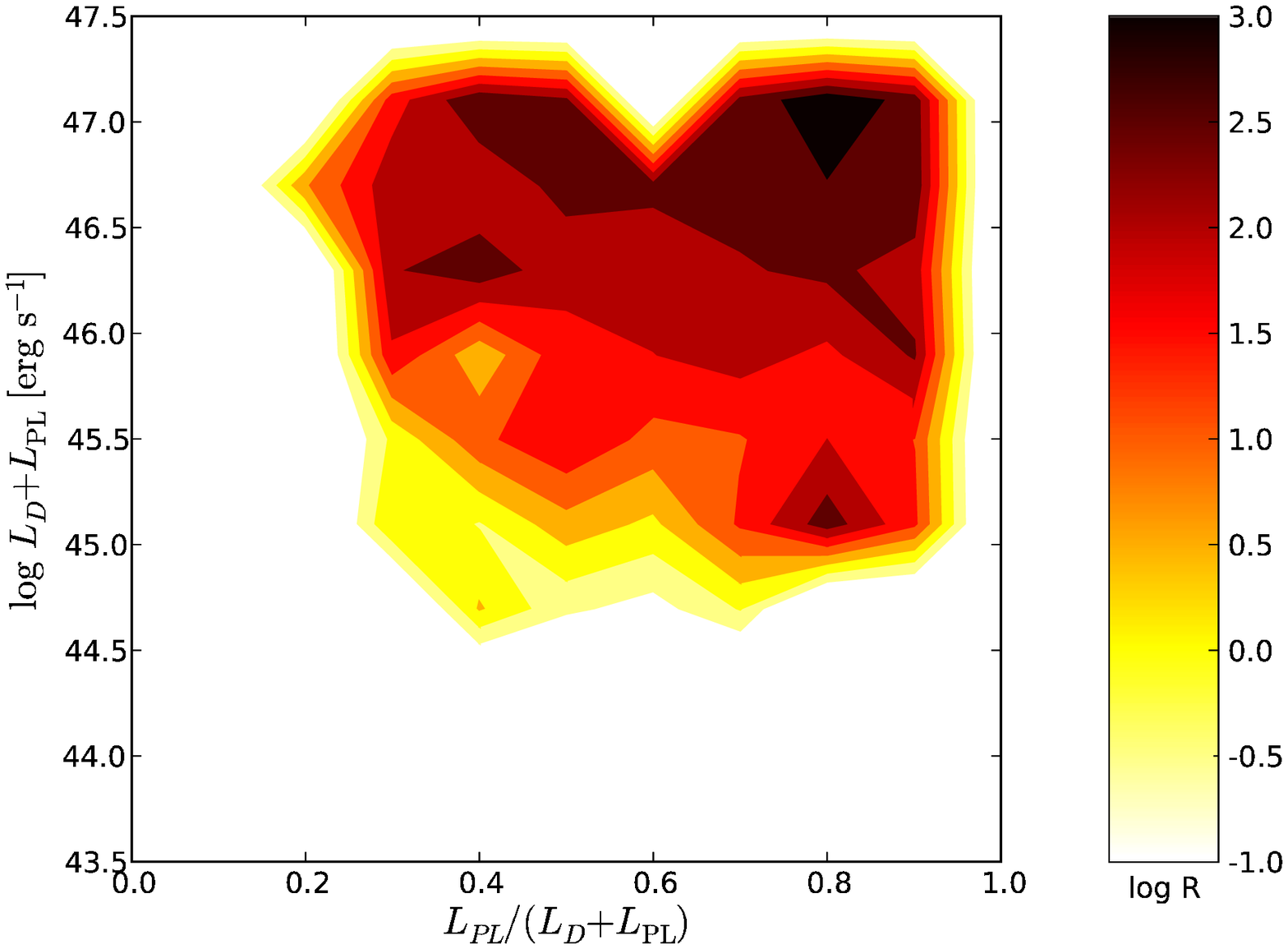}}
\resizebox{8.7cm}{!}{\includegraphics{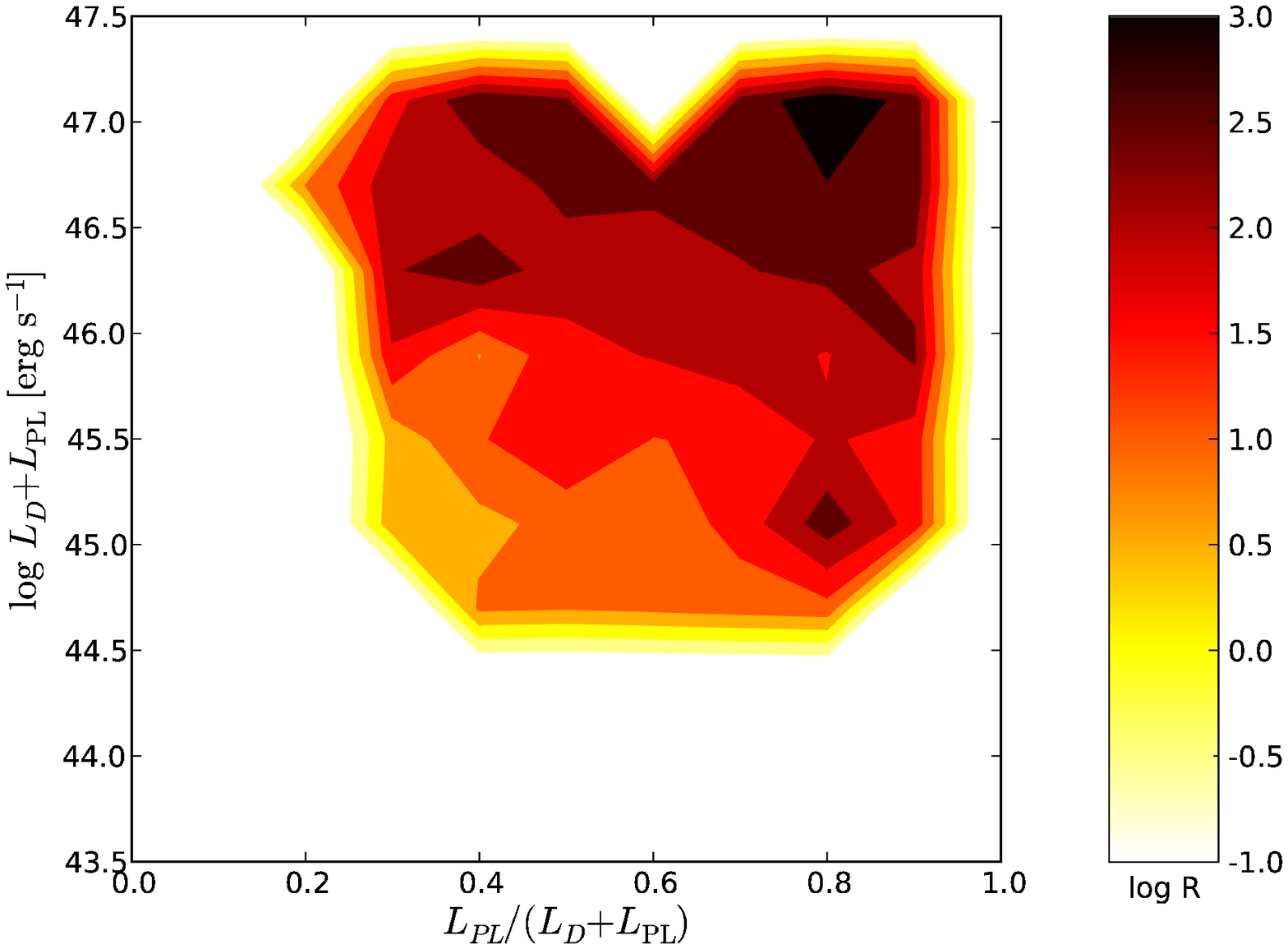}}
\caption{Left: DFLD of 4963 SDSS quasars with ROSAT detections in
  the redshift range $0.2\leq z \leq 2.5$, showing the lower limit to
  the average radio loudness $R$ as calculated from FIRST data. The
  contour levels are separated by 0.5 dex in $R$. Right: Upper
  limit to the average $R$. The contour levels are the same as for the
  lower limit.}
\label{fiMapR}
\end{figure*}

To construct a DFLD for AGN we need a sample of AGN with well measured
optical ($\Ld$), X-ray (\Lpl) and radio fluxes.

\subsubsection{SDSS Quasars}
\label{s:method.sample.sdssqso}

As our main sample, we begin with the lists of all objects from the
spectroscopic database\footnote{\url{http://cas.sdss.org/astro}} of
SDSS DR5 \citetext{J.\ Adelman-McCarthy et al., \emph{in prep.};
  \citealp{AdelmanAguerosAllam2006}} that have been identified as a
quasar with high confidence. The sample includes all objects from the
statistically complete quasar survey \citep{RichardsFanNewberg2002} as
well as quasar spectra selected in other ways, e.g., candidate optical
counterparts of known X-ray sources \citep{AndersonVogesMargon2003}.
Given the criteria used by the SDSS to classify objects as quasars,
all objects are Type~1 quasars with strong broad emission lines (and
potentially broad absorption lines).  Appendix~\ref{s:sql_app} gives
the SQL query necessary to reproduce our initial sample.  For X-ray
and radio measurements, we use data from the ROSAT All-Sky Survey
\citep[RASS, energy band 0.1--2.4~keV;][]{VogesAschenbachBoller1999}
and the Very Large Array (VLA) Faint Images of the Radio Sky at Twenty
centimeters (FIRST) survey \citep[observing frequency
  1.4~GHz;]{WhiteBeckerHelfand1997} as given in the SDSS database. The
radio fluxes are intended to be a measure of the core radio flux of
our sources (see \S\ref{s:disc.impact.size}).  We only include quasars
in the redshift range $0.2 \leq z \leq 2.5$.  Below a redshift of 0.2,
not many quasars are found, and spectra of low-luminosity AGN are
often dominated by galaxy light, making them more likely to be
classified as galaxy by the SDSS pipeline.  Redshifts above 2.5
contribute only a small fraction of the total number of SDSS quasars,
and not many of them have a ROSAT detection at such high redshifts. In
the adopted redshift range, there are a total of 64248 objects of
which 5174 have a radio detection, 4963 have an X-ray detection, and
730 have detections in both bands.

We apply a $K$-correction to the flux measurements in all three
wavebands in order to remove apparent luminosity differences that are
caused by observing objects at different redshifts, and therefore at
different rest wavelengths.  The $K$-correction is conventionally
defined as $m_{\mathrm{intrinsic}} = m_{\mathrm{observed}} - K(z)$
\citep[see][e.g.]{HoggBaldryBlanton2002}.  For the SDSS quasars, we
use reddening-corrected $i$-band PSF magnitudes (identical to those
used to flux-limit the SDSS main quasar sample) and apply the $i$-band
$K$-correction given by \citet[Table 4]{RichardsStraussFan2006}.
These authors remove the average contribution of emission lines as
function of redshift to obtain a pure continuum luminosity with an
assumed power-law spectrum $f_\nu \propto \nu^{-0.5}$ and define the
correction to be 0 at $z=2$.  The continuum $K$-correction is easy to
compute since the SDSS magnitude system is (nearly\footnote{See the
  discussion at
  \url{http://www.sdss.org/dr5/algorithms/ fluxcal.html}}) 
an AB system \citep[defined as $\mathrm{mag} = -2.5 \log
  (f_\nu/3631\,$Jy$)$ for a source with constant
  $f_\nu$][]{OkeGunn1983,AbazajianAdelmanAgueros2003}; the correction
is given by $K(z) = -2.5(1+\alpha)\,\log(1+z)$ for $f_\nu \propto
\nu^{\alpha}$.  We convert the $K$-correction to $M_i(z=2)$ back to the conventional
$M_i(z=0)$ by adding the offset 0.596 appropriate for an object at
redshift $z=2$ with the assumed power-law spectrum with $\alpha=-0.5$
\citep[see][eqn.\ {[1]}]{RichardsStraussFan2006}. We then transform to the
$B$-band using $B-i = 0.3$ as appropriate for the
same power-law spectrum.  Thus, the $B$-band magnitude is obtained
from the SDSS $i$-band PSF magnitude, galactic extinction $A_i$ as
given in the SDSS database, and the combined emission-line and
continuum $K$-correction $K_\mathrm{opt}$ as
\begin{equation}
\label{eq:optKcorr}
B = i + 0.3 - A_i - K_\mathrm{opt} + 0.596.
\end{equation}
However, we do not apply any correction for contributions from host
galaxy starlight. This correction is most important for the lowest
quasar luminosities and for the lowest-redshift objects.  We discuss
the impact of this simplification in \S\ref{s:disc.impact.beaming}.

For the $K$-correction of the radio fluxes we assume optically thin
spectra with a spectral shape $f_\nu \propto \nu^{-0.5}$, i.e., we
assume the same spectral shape for the radio and optical continua.
For the X-ray data, we assume a photon index $\Gamma=2$, i.e., a
spectral shape $f_\nu \propto \nu^{-1}$ \citep[found to be appropriate
for both radio-loud and radio-quiet quasars
by][]{GalbiatiCaccianigaMaccacaro2005} which has 0 $K$-correction
because the effects of bandwidth narrowing and observing a steep
spectrum at lower rest-frame frequencies cancel.

As the SDSS quasar sample includes objects with a large range of
black-hole masses, it is in principle necessary to apply a mass
correction to the observed luminosities.  Accreting black holes that
are observed as a quasar all have similar Eddington ratios, in the
range 0.05--1 \citep[e.g.,][]{Jester2005}. Hence, to zeroth order, the
luminosity of a quasar is expected to scale as the Eddington
luminosity, i.e., linearly with black hole mass.  However, the black
hole masses computed for 12245 quasars from SDSS DR1 by
\citet{McLureDunlop2004} have a mean of $(4.75\pm0.44)\times
10^8$\,M$_{\sun}$, i.e., the intrinsic scatter in the black hole mass
of quasars accessible to the SDSS spectroscopic survey is comparable
to the scatter in the black-hole mass determination \citep[estimated
  to be as large as 0.5--0.6 dex by][]{Vestergaard2004}.  Hence, the
black-hole masses of all SDSS quasars are consistent with being equal
to the mean and we apply no mass scaling.  This needs to be borne in
mind when comparing this quasar sample to AGN samples with
substantially different black hole masses.

We use a cosmology with $\Omega_M=0.3, \Omega_\Lambda=0.7,
H_0=71$~km\,s$^{-1}$\,Mpc$^{-1}$.

\subsubsection{Low-luminosity sample}

While our main results will be based on the SDSS quasars, we construct
a low-luminosity sample based on the sample of
\citet{HoFilippenkoSargent1997a} as a comparison. We have obtained
X-ray fluxes for this sample from \citet{TerashimaWilson2003} and the
following surveys in order of preference: The Chandra v3 pipeline
(\citealt{PtakGriffiths2003}\footnote{\url{http://www.xassist.org}}),
the XMM serendipitous X-ray survey \citep{BarconsCarreraWatson2002},
and the ROSAT HRI pointed catalog \citep{Rosat2000}. Radio fluxes for
the sample have been taken from \citet{NagarFalckeWilson2005}. For
low-luminosity AGN, the optical emission is often dominated by galaxy
light. Therefore, we use a correlation between H$\beta$ and the B-band
magnitude \citep{HoPeng2001}:
\begin{equation}
\log L_{\mathrm{H}\beta} = -0.34 M_B + 35.1
\end{equation}
to compute the optical absolute magnitude.  This correlation seems to
hold for quasars as well as Seyfert galaxies. We use the correlation
also for our low-ionization nuclear emission region (LINER) objects
even though the correlation has only been shown to hold for Seyfert objects
and PG quasars.

The average BH mass of the SDSS quasars is significantly larger than
that of the nearby LLAGN. Thus, for LLAGN it is important to consider
their BH masses.  The BH masses of the LLAGN sample are calculated
from the M-$\sigma$ relation \citep{MerrittFerrarese2001} using
velocity dispersions from the Hypercat catalog
\citep{PrugnielZasovBusarello1998}.  We scale the luminosities
linearly with mass to the average mass of the SDSS quasars.

\subsection{Calculating the component luminosities}

To compute the accretion-disc luminosity, we use the relation between
the absolute optical $B$-band magnitude and the approximate disc
luminosity given by \citet{FalckeMalkanBiermann1995}:
\begin{equation}
\log \Ld = -0.4 M_B +35.9
\end{equation}
This disc luminosity measure has a scatter of 0.29 dex.

The power-law component $\Lpl$ is estimated from the X-ray
luminosity. To obtain the power-law luminosity, we extrapolate the
measured ROSAT counts to a flux in the 0.5-10 keV band using a photon
index of $\Gamma = 2$. For the hydrogen column density we use an average 
Galactic column density of $n_H \sim 1 \times 10^{21}$ cm$^{-2}$ \citep[e.g.,][]{DickeyLockman1990}. 
The hydrogen column density in the Quasar is usually negligible of broad line Quasars \citep[for the low absorption of broad line objects see, e.g.,][]{RichardsHallVandenBerk2003}, additionally, 
the large redshift reduces its effect. 
This yields a conversion factor of $1\,
\mathrm{cps}\  \mathrel{\widehat{=}} 3.1\times 10^{-11}$ erg s$^{-1}$ cm$^{-2}$. Where
Chandra observations are available, we can measure the flux in this
energy range directly. The luminosity of the power-law component is
chosen to be
\begin{equation}
\Lpl = 3 L_{0.5-10\mbox{keV}}. 
\end{equation}
This ``bolometric correction'' of a factor 3 is fairly arbitrary and
was chosen to distribute the quasars uniformly over the $x$-axis of
the DFLD.  For the purpose of our analysis, it does not have to
correspond directly to the real bolometric correction of the quasar,
but is probably not far off.  However, a different value creates only
a slightly different mapping of the sources on the defined axis. This
is similar to the definition of the hardness ratio in the HIDs, which
is also arbitrary and chosen to give a good distribution of the
sources in the HID.  Combining these ``bolometric'' and the
$K$-corrections from \S\ref{s:method.sample.sdssqso} into the factors
$c_{\mathrm{opt}}$ and $c_{\mathrm{X}}$, our non-thermal fraction is
related to the observed fluxes in both wavebands by $1/(1+
c_{\mathrm{opt}} f_{\mathrm{opt}}/[c_{\mathrm{X}}f_{\mathrm{X}}])$.

To compare the radio loudness of the AGN as function of their position
in the DFLD, we also calculate the parameter $R$, which is defined as
the ratio of the radio flux at 1.4 GHz divided by optical $B$-band
flux. For inclusion in this diagram, we only accept sources which have
optical and X-ray fluxes, but we include sources with only an upper
limit for $R$. Including the expressions for our $K$-correction terms
from \S\ref{s:method.sample.sdssqso}, $R$ is related to fluxes from
the catalogues by the expression
\begin{equation}
\log R = 0.4 (B-t)
\end{equation}
\citep{IvezicMenouKnapp2002}, where $t$ is the $K$-corrected 1.4 GHz
radio AB magnitude given by
$t=-2.5[\log(f_{\mathrm{FIRST}}/3631\,\mathrm{Jy})+0.5\log(1+z)]$ and
$B$ is the $K$-corrected $B$-band magnitude from
eqn.\ (\ref{eq:optKcorr}).

\section{Results}
\label{s:results}

We can now use our samples of AGN to construct DFLDs, i.e., the
$y$-axis of this diagram denotes $\Ld+\Lpl$, while on the $x$-axis we
plot $\frac{\Lpl}{\Ld+\Lpl}$. For every source, we can calculate the
position in the DFLD and its radio loudness $R$, or an upper limit to
$R$. To study the dependence of radio loudness on location in the
DFLD, we compute the average radio loudness of AGN at each
position. We do so by binning the data in $10 \times 10 $ cells of the
DFLD and calculating the arithmetic mean of $R$ for each cell. We only
show those cells in the diagram that contain more than 10 sources. We
treat the upper limits to $R$ in two ways: First we set $R$ to zero
for all sources without a radio detection. This average $\Rlo$ is a
lower limit to the real average of $R$. Secondly, we set $R$ to the
upper limit of the corresponding radio measurement. This average,
$\Rhi$, is an upper limit to the real average $R$. Thus, the real
average $R$ has to lie between these two values. We will use these
values in our figures to detect selection effects caused by the FIRST
survey's radio flux limit.

\subsection{SDSS Quasars}
\begin{figure}
\resizebox{8.7cm}{!}{\includegraphics{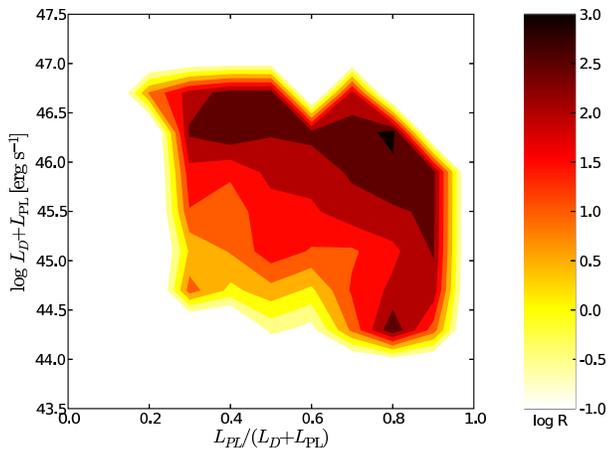}}
\caption{Modified DFLD showing the average radio-loudness. The y-axis
gives the optical disc luminosity $L_D$ instead of the sum of the disc
and the power law luminosity in the previous maps. The most radio-loud
sources are those with a large power law component in the spectrum.}
\label{fiMapLD}
\end{figure}

In Fig.~\ref{fiMapR} we show both upper and lower limit of the average
$R$ for our 4963 SDSS quasars with ROSAT detections.  The average $R$
value is largest in the upper right corner and declines gradually
towards the lower left corner. The average $\log R$ in the top right
corner is 2.6 dex higher than in the bottom left corner, which
corresponds to a factor 398 in $R$. The gradient is seen both in the
upper and lower limit of $R$ and the difference between both maps is
very small. Thus, we conclude that this effect is also found in the
real averaged $R$. The effect can be described as combination of two
trends: First, the most radio-loud objects for a given disc luminosity
(i.e., optical luminosity) are among those sources that are the most
X-ray bright (see Fig.~\ref{fiMapLD}). Additionally, brighter objects
tend to be more radio-loud for a given disc fraction (a fixed
$x$-value in the DFLD). The first effect is already known as
radio-loud objects are known to be more X-ray bright (e.g.,
\citealt{ElvisWilkesMcDowell1994,ShenWhiteMo2006}). 

\begin{figure}
\resizebox{8.7cm}{!}{\includegraphics{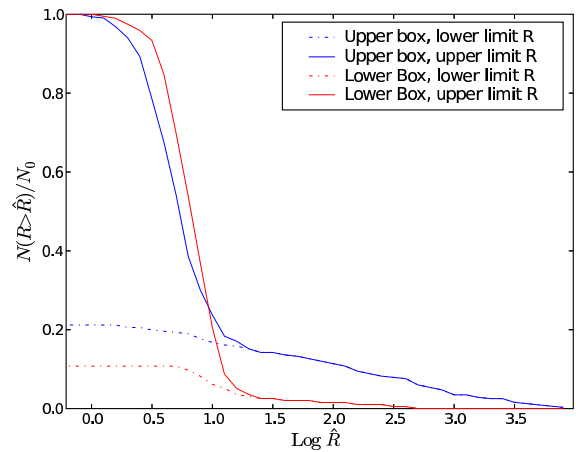}}
\caption{Log $N$-log $R$ diagram for two rectangular areas in the DFLD
diagram.  The position and extent of the upper and lower box is given
in the text. For both areas we first find all objects with a position
on the DFLD in that area and construct log $N$-log $R$ diagrams for
both areas individually. Solid lines indicate upper limits to the
average $R$ while dashed lines represent its lower limit. }
\label{fiLogNLogS}
\end{figure}

As the average has been computed separately for every cell, two
different cells are statistically independent. Thus, even though the
average in each cell does not have to be significant, the fact that
the map is smooth and has a well-defined gradient in $R$ indicates
that this effect is not a statistical artifact. To verify this
observation mathematically, we have performed a Kolmogorov-Smirnov
test of the differences between the distributions of radio loudness in
two well-separated areas of the DFLD. We compare all sources within a
``high-luminosity'' area given by $0.3 \leq x \leq 0.5$ and $46 \leq y
\leq 47$ to those in a ``low-luminosity'' area given by $0.3 \leq x
\leq 0.5$ and $44 \leq y \leq 45.5$. In Fig.~\ref{fiLogNLogS} we show
the Log $N$ -- Log $R$ diagrams for both areas. Already visual
inspection suggests that the ``high-luminosity'' area contains a
significantly larger fraction of radio-loud objects. This is confirmed
by the Kolmogorov-Smirnov test, which rules out that the two
underlying probability distributions are identical with a confidence
level of 99.994 \% ($D = 0.19$). We also verified that the
radio-loudness distribution is different for areas with similar
luminosity but different disc-to-power law ratios.

\begin{figure*}
\resizebox{8.7cm}{!}{\includegraphics{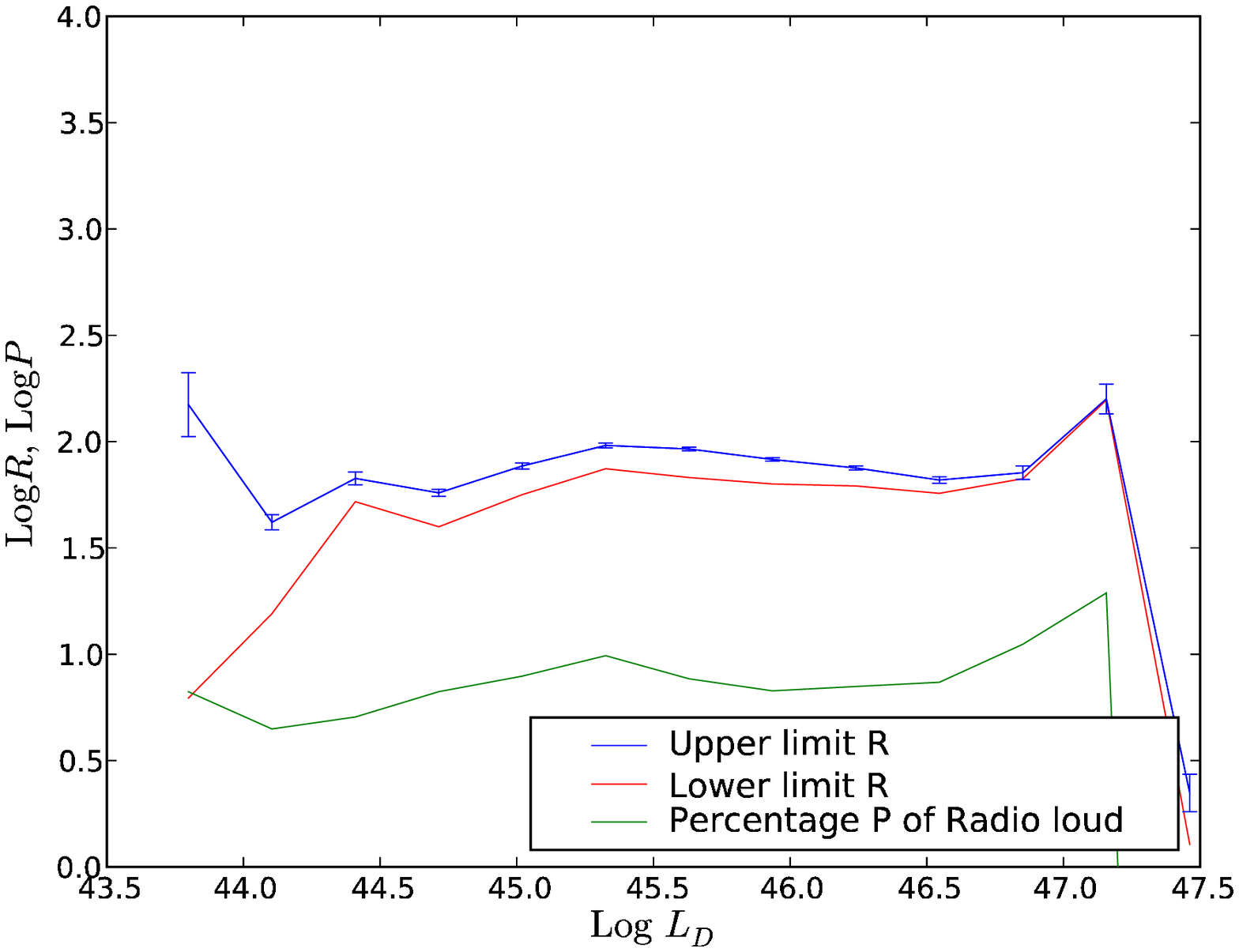}}
\resizebox{8.7cm}{!}{\includegraphics{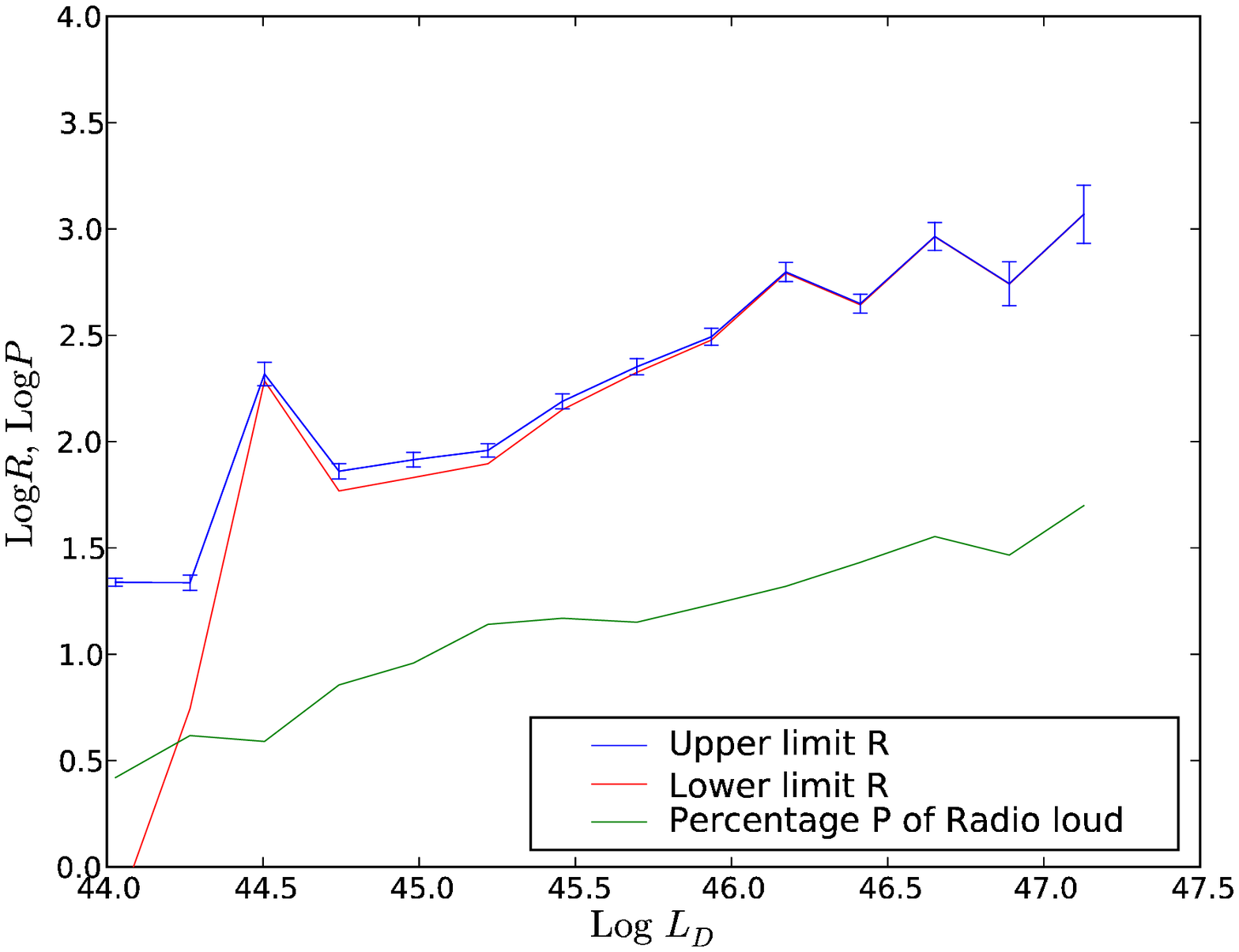}}
\caption{Limits to average radio loudness and percentage of radio-loud
  sources ($R>10$) as function of disc luminosity. Left: all 64248
  SDSS quasars in the redshift range $0.2 \leq z \leq 2.5$. Right:
  only the 4963 X-ray detected sources.  The error bars give the
  uncertainty of the mean in each of the bins.}
\label{fiProjectionR}
\end{figure*}

\begin{figure}
\resizebox{8.7cm}{!}{\includegraphics{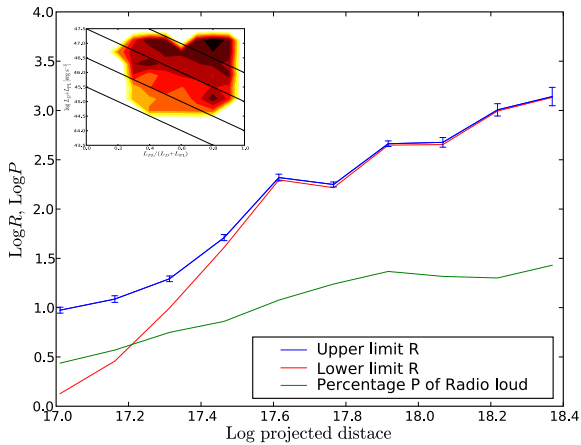}}
\caption{Projection parallel to the black lines in the DFLD shown in
  the inset (with the chosen scaling of the coordinate axes, the line
  \emph{onto} which objects are projected is not perpendicular to
  these lines). We project on to the line defined by
  $\bmath{\lambda}=(0.93\,\bmath{e}_x +0.36\,\bmath{e}_y)$. This
  projection includes only sources with measured X-ray fluxes.}
\label{fiProjectionR2}
\end{figure}

To verify that the X-ray emission does have an effect on the radio
loudness, we plot the average radio loudness as a function of the disc
luminosity (i.e., the optical luminosity) in the left-hand panel of
Fig.~\ref{fiProjectionR}, for all 64248 SDSS objects, independently of
their X-ray detection. The average $R$ has been estimated
independently in 20 logarithmic luminosity bins. This plot of the
radio loudness as a function of luminosity can be compared to the
right-hand panel of the same figure, where we show the dependence of
the radio loudness on the disc luminosity for the 4963 X-ray detected
sources only.  While there is the clear trend visible in the X-ray
detected quasars that brighter objects are more radio-loud, only
marginal effects can be seen in the full sample. We show the upper and
lower limit to the radio loudness as well as the fraction of
radio-loud objects as defined by $R>10$. The trend can be seen in all
three quantities. We can also find a clear trend in the radio-loudness
if we project the DFLD onto an axis roughly corresponding to the
diagonal. In Fig.~\ref{fiProjectionR2} we show such a projection for
the SDSS sample \citep[compare][Fig.\ 11]{ShenWhiteMo2006}.

\subsection{Hard-state objects: LLAGN}\label{seLLAGN}

So far, we have only considered the SDSS quasars. All quasars are
thought to be strongly accreting objects with a standard geometrically
thin, optically thick disc. To obtain information about the shape of
the DFLD at lower luminosities, which may be in a different,
radiatively inefficient accretion mode, we now include the sample of
\citet{HoFilippenkoSargent1997a}. Due to the small number of sources
we now show every cell for this sample, even if it includes only one
source. Hence, for the LLAGN, the plotted ``mean'' $R$ may just be a
single measurement that could be substantially different from the true
mean $R$.

\begin{figure}
\resizebox{8.7cm}{!}{\includegraphics{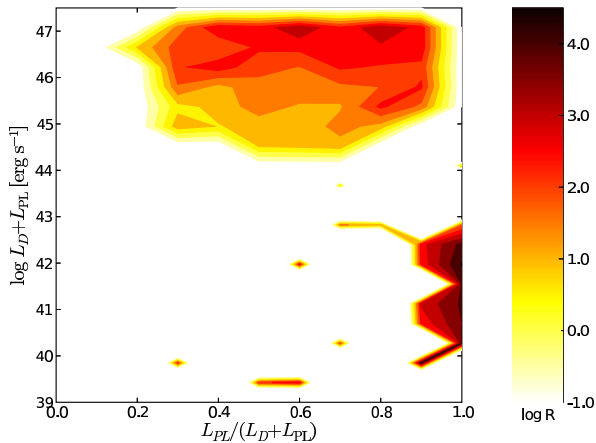}}
\caption{DFLD showing the average radio loudness for SDSS quasars and
  LLAGN from the \protect\citet{Ho1999} sample. Note that the 
gap between LLAGN and Quasars is an artefact of our sample selection.}

\label{fiHoSample}
\end{figure}
In Fig.~\ref{fiHoSample} we show the joint DFLD for the SDSS quasars
and the LLAGN sample. Not surprisingly, the LLAGN lie at lower
luminosities than the SDSS quasars. However, except for a few outliers
they all lie at the right sight of the DFLD.  This is also not
surprising, as \citet{Ho1999} already noted that LLAGN are less
optically bright for a given X-ray luminosity than strongly accreting
AGN (Seyfert galaxies and quasars). Furthermore, the LLAGN in our DFLD
seem to be more radio-loud on average than SDSS quasars. As mentioned,
we have to plot every cell for the LLAGN even if it only contains one
object, so the mean value of $R$ for any given cell may not be close
to the true mean. But already \citet{HoPeng2001} have shown that the
average low-luminosity Seyfert galaxy is more radio-loud than the
average ``radio-loud'' PG quasar if only the nuclear luminosity is
considered (i.e., removing the contribution of starlight).  Thus, we
find that LLAGN populate the bottom right-hand side of the DFLD and
are --- on average --- more radio-loud than the SDSS quasars.

\section{Critical analysis of selection effects and astrophysical
biases}
\label{s:critical}

In this section, we consider whether the shape of the distribution of
SDSS quasars in the DFLD and their average radio loudness could be
caused entirely due to selection effects, or underlying correlations
between the observables and other parameters.  We consider in detail
the selection effects of flux-limited samples (\S\ref{s:crit.sel}),
black-hole mass scalings (\S\ref{s:crit.Mbh}), orientation, beaming
and host-galaxy light (\S\ref{s:disc.impact.beaming}) and the
size-luminosity evolution of radio
sources(\S\ref{s:disc.impact.size}). Based on our analysis, none of
these effects can produce the entire signal we report.

\subsection{Selection effects}
\label{s:crit.sel}

Our main sample is basically a single quasar survey with a fairly
bright flux limit --- most quasars in our sample are drawn from that
part of the quasar sample limited by $i<19.1$. Even though we include
several other SDSS target categories, there is an implicit flux limit
$i \la 20$ imposed by the signal-to-noise that can be achieved in the
exposure time of the spectroscopic observations in the SDSS.  As the
optical luminosity function of quasars is rather steep, this means
that the dominant selection effect is that we only have a rather
restricted dynamic range in quasar luminosity at fixed redshift, of
order 10--30 (the SDSS spectroscopic survey has a bright limit $i>15$
and substantial numbers of objects are only found at $i\ga16$).  Thus,
only the most luminous object at any given redshift are found, and the
variation in luminosity is largely due to a variation in redshift.
Thus, on average there is a clear trend that the redshift of
the lowest luminosity sources is small ($z=0.2$) while the highest
luminosities are found at large distances ($z = 2.5$).  

\begin{figure}
\resizebox{8.7cm}{!}{\includegraphics{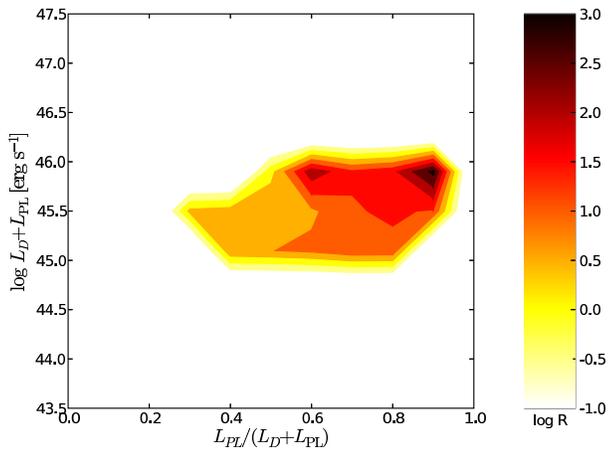}}
\caption{ DFLD of the redshift slice $0.3 \leq z \leq 0.4$. In this redshift range our sample contains 676 sources with optical and X-ray detections of which 70 also have radio detections. Due to the small number of available sources we show all bins with more than 5 sources. Similar to the DFLD of the full sample sources in the upper right corner have large average $R$ values, while those in the lower left are less radio-loud. }
\label{fiSlice}
\end{figure}

To assess the impact of this well-known degeneracy between redshift
and luminosity in flux-limited surveys, we have verified that the
effect is visible in small redshift slices, e.g., in the subsample
with redshifts between $z=0.3$ and $z=0.4$ as shown in Fig.~\ref{fiSlice}. Furthermore, we have
verified that the $K$-correction is not creating the observed trends
in the DFLD. Even if we assume a flat radio spectrum or do not
$K$-correct at all, we find the observed effect. The difference in
radio loudness between the average SDSS quasar at low redshift
($z\approx 0.2 $ ) and those at large redshifts ($z\approx 2.5$) is
0.6 in $\log R$.  This difference is probably mainly due to a
difference in accretion rates.  However, even if this is purely due to
evolution, we note that the observed range in $R$ in the DFLD is
around 2.5 dex, so evolution would only be a minor effect.  This
suggests that the radio loudness gradient in the DFLD is not an
artifact of redshift evolution, but we cannot rule out that evolution
plays a minor role.  It clearly will be desirable to construct a DFLD
with a sample that is not affected by evolution (i.e., one with a
narrow redshift distribution).

We have shown the average radio loudness as a function of the position
in the DFLD in two versions: as an upper limit and as a lower limit to
the real radio loudness (i.e., the value we would observe with a radio
telescope of infinite sensitivity). More sensitive radio observations
would increase the lower limits while they would decrease the upper
limit map. As there are only minor differences between both maps, we
conclude that we our result is not affected strongly by the radio
detection limit. 

\begin{figure}
\resizebox{8.7cm}{!}{\includegraphics{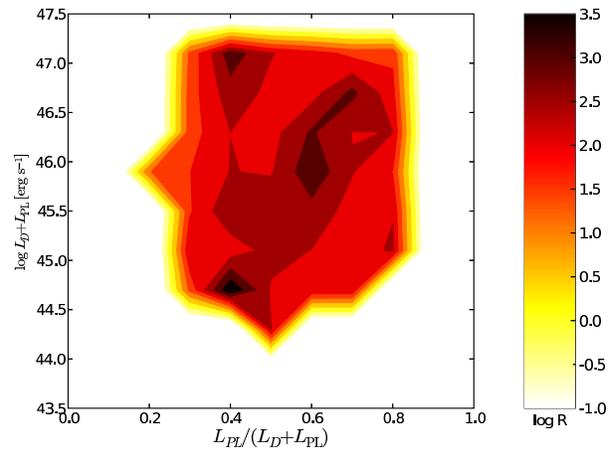}}
\caption{Monte-Carlo simulation of the radio loudness as a function of
the position on the DFLD, assuming no correlation between radio,
optical and X-ray luminosities.  We assume simple power-law luminosity
functions and give the simulated sample the same redshift distribution
as our observed sample.}
\label{fiMonteCarlo}
\end{figure}

As an additional test, we have performed a Monte-Carlo simulation to
check whether the trends observed in the DFLD for the SDSS quasars are
merely due to selection effects arising from the use of flux-limited
surveys. As null hypothesis, we assume that radio, optical and X-ray
fluxes of all objects are uncorrelated, so that the conditional
luminosity function is simply the product of the individual luminosity
functions: $\Phi(\Lrad,\Lopt,\Lx) =
\Phi_1(\Lrad)\,\Phi_2(\Lopt)\,\Phi_3(\Lx)$. At the high luminosities
that are accessible to our three surveys, all three luminosity
functions are well-described by simple power laws $\Phi(L) \propto
L^{\beta_\mathrm{R,O,X}}$. For the radio luminosity function we use $
\beta_\mathrm{R} = 0.78 $ \citep[][]{NagarFalckeWilson2005},
for the X-rays $\beta_\mathrm{X} = 2.34$ \citep{UedaAkiyamaOhta2003}
and for the optical luminosity function $\beta_\mathrm{O} = 2.95$
\citep{RichardsStraussFan2006}. For this simple test we assume that
the power law indices do not change with redshift. The number of
sources with radio, optical and X-ray detections at a given redshift
is fixed to the observed value. This incorporates density or
luminosity evolution with redshift, as well as the redshift-dependent
selection efficiency of the SDSS quasar survey. The resulting DFLD of
the simulated sample is shown in Fig.~\ref{fiMonteCarlo}. The trend
visible in the observed DFLD (Fig.~\ref{fiMapR}), i.e., an increase in
the mean radio loudness from the bottom left to the top right, is not
observed.  Futhermore, we have repeated this luminosity function test
with a population of artificially beamed steep-spectrum sources: a
population of steep-spectrum sources with
$\beta_\mathrm{R} = 2.31$ \citep{WillottRawlingsBlundell2001}, 10\% of
which have their radio flux increased by a factor 100, but still
without correlation to optical or X-ray properties.  This simulation
cannot reproduce our findings, either.

\subsection{Black-hole mass}
\label{s:crit.Mbh}

We are not able to correct the DFLD of our SDSS quasars for the BH
mass as we do not know the mass for most of our objects. Even though
most quasars in the SDSS have masses around $8.7 \times 10^8 M_\odot$
\citep{McLureDunlop2004}, more luminous objects are more likely to
have a more massive black hole. Thus, objects in the lower-luminosity
part of the diagram will have a lower mass on average, while
high-luminosity objects will have larger BHs.  It has been suggested
that the radio power of AGN correlates with BH mass
\citep{FranceschiniVercelloneFabian1998,LacyLaurent-MuehleisenRidgway2001,JarvisMcLure2002}. Also
the fundamental plane of black-hole activity
\citep{MerloniHeinzdiMatteo2003,FalckeKoerdingMarkoff2004} suggests a
weak mass dependence as the radio luminosity depends nonlinearly on
the accretion rate (not scaled to Eddington units).  However, other
researchers have found no evidence for correlations between radio
loudness or radio power and black-hole mass (see
\citealt{Ho2002,WooUrry2002b}, e.g.), as selection effects severely
affect most of the studied samples. Thus, it is important to explore
whether a mass dependence of the radio loudness can modify our
findings.

In the left-hand panel of Fig.~\ref{fiProjectionR}, we plot the
average radio loudness against disc luminosity, i.e., the optical
luminosity. We do not find a strong correlation between the two as
long as we consider all SDSS quasars. Only when we exclude X-ray
non-detections, a clearer trend is found. But also in the first plot
the most luminous objects should have --- on average --- larger BH
masses than the faint objects. Thus, a strong mass effect is not seen
for all SDSS quasars with radio detections. Furthermore, it is
well-known that radio-loud objects are more X-ray bright (e.g.,
\citealt{ElvisWilkesMcDowell1994}). This supports the idea that the
observed dependence of radio loudness on position in the DFLD is not
purely due to different black hole masses.

Jet models often assume a linear coupling between the accretion power
and the power injected into the jet
\citep[e.g.,][]{FalckeBiermann1995,Meier2001}. This suggests that
larger black holes should have larger radio luminosities as the
physical accretion rate is higher for a fixed Eddington ratio. In our
DFLD we show the radio loudness, i.e. the ratio of the radio flux and
the flux in the $B$-band. A constant $R$ indicates that the radio flux
scales linearly with optical flux and probably the accretion
rate. However, it may be that the radio luminosity scales with a
higher power of the accretion rate. E.g., for the radio core, the
radio luminosity scales roughly with $L_{Rad} \propto \dot{M}^{1.4}$
if the jet power is a constant fraction of the accretion rate
\citep{BlandfordKonigl1979}. This scaling is further supported by the
``fundamental plane of accreting black holes''
\citep{FalckeKoerdingMarkoff2004,MerloniHeinzdiMatteo2003,KoerdingFenderMigliari2006}. We have
verified that this effect does not change the plot of the
radio-loudness in the DFLD. The figure looks similar if we plot
$L_{Rad}/\Ld^{1.4}$ (this assumes that the disc luminosity is a good
tracer of $\dot{M}$). Thus, this dependence of the radio luminosity on
accretion rate cannot reproduce the observed diagrams.

It has been suggested that instead of a \emph{correlation} between $R$
and black-hole mass, there is a hard switch in radio loudness for
source with BH masses $M > 10^8 M_\odot$
\citep[e.g.,][]{Laor2000,McLureJarvis2004}.  The Eddington limit for a
$10^8 M_\odot$ BH is around $10^{46}$ erg s$^{-1}$, but our DFLD still
shows a gradient in radio loudness above this luminosity.  This
gradient can therefore not be solely due to a single switch. Thus, if
the BH mass is the dominant effect in the DFLD, there must be an
additional mass dependence besides the suggested switch around $M >
10^8 M_\odot$.

As a ``worst-case scenario'', let us assume that all objects have a
fixed Eddington ratio. In this case, the luminosity range of 2 orders
of magnitude is due to 2 orders of magnitude difference in the
black-hole mass. The highest and lowest average values of $R$ found in
our DFLD differ by a factor $10^{2.5}$. If the differences in radio
loudness were purely due to a mass dependence, this would suggest that
$\Lrad \propto M^{2.3}$. Furthermore, if this effect was to explain
the entire radio-loudness distribution on the DFLD, the most massive
black holes would have to be the most X-ray bright objects (i.e., lie
on the right-hand side of the diagram). This does not seem to be the
case (e.g., \citealt{WooUrry2005}). Since the required mass scaling
seems to be rather large (compare Fig. 2 in
\citealt{McLureJarvis2004}) and the needed X-ray to optical properties
are not found, we consider it unlikely that the radio loudness
distribution in the DFLD arises purely due to a mass scaling of the
radio loudness.

\subsection{Orientation, beaming, and host galaxy light}
\label{s:disc.impact.beaming}

The orientation of a quasar's accretion disc and jet with respect to
the observer can play an important role for the observed luminosities
(see e.g., \citealt{UrryPadovani1995}). The main effects are
relativistic beaming of the emission originating from the jet and
obscuration of disc and broad-line emission by a torus. All our SDSS
quasars are of Type~1, so broad lines are visible in all objects. It
follows that we select only sources that are not strongly obscured
\citep[see also the discussion of the likely fraction of reddened
quasars in the SDSS
by][]{RichardsHallVandenBerk2003,HopkinsStraussHall2004}. This is
especially important as we use soft X-ray data from ROSAT, which is
very sensitive to absorption. Due to our selection of broad-line
objects, we ensure that the emission lines are not swamped by the
featureless continuum of the relativistic jet (a BL Lac object would
not be classified as a broad-line SDSS quasar). Hence, we are selecting
against strongly boosted sources.

Thus, we can safely assume that the optical emission is mostly due to
the accretion disc. Radio emission, however, is known to be strongly
affected by beaming, because it arises from a relativistic jet.  If
radio emission is the only beamed component in the SED, any
correlation between the strength of the beaming and the position of
the source in the DFLD supports our statement that the jet properties
of AGN depend on both the optical and X-ray properties of the source.

In case that some or all of the X-ray emission also originates from
the jet, the X-ray bright sources on the right side of the DFLD will
preferentially be stronger beamed. In unified schemes for AGN
\citep[e.g.,][]{UrryPadovani1995}, flat-spectrum radio quasars (FSRQ)
are objects whose radio and optical continuum emission is enhanced by
relativistally beamed emission from a jet; hence, part of the X-ray
emission will also be enhanced by beaming.  However, the ``blazar
sequence'' \citep{FossatiMaraschiCelotti1998} appears to indicate that
beamed X-ray emission is progressively less important in objects with
higher optical (accretion disc) luminosity \citep[although a number of
  recent authors have questioned the reality of the blazar sequence;
  e.g.,][]{AntonBrowne2005,LandtPerlmanPadovani2006,NieppolaTornikoskiValtaoja2006}.
Based on correlations between X-ray, radio and optical fluxes, a
  number of authors concluded already in the \emph{Einstein} era that
  there has to be a beamed component of X-ray emission in sources with
  beamed radio emission
  \citep{KembhaviFeigelsonSingh1986,BrowneMurphy1987,Shastri1991,ShastriWilkesElvis1993,JacksonBrowneWarwick1993,Kembhavi1993}.
Later, \citet{BakerHunsteadBrinkmann1995} found a beaming-induced
correlation between radio and X-ray fluxes both for steep-spectrum and
flat-spectrum quasars, while both
\citet{GalbiatiCaccianigaMaccacaro2005} and
\citet{GrandiMalagutiFiocchi2006} report evidence for a beamed
contribution in FSRQ, but not in other radio-loud quasars.  On the
other hand, \citet{EvansWorrallHardcastle2006} found a contribution to
the X-ray emission of high-power radio galaxies and quasars from
\emph{both} the accretion flow (i.e., the disc) and the jet,
indicating that the X-ray emission is not \emph{purely} due to a
beamed component.

To consider whether beaming is important in the optical, we begin with
the observation that the H$\alpha$ equivalent width of \emph{all}
types of AGN is essentially constant
\citep[Fig.~11.6]{MyOsterbrock1989} and independent of the radio
properties (see also \citealp{BorosonGreen1992}, \S\,4.2; this
statement is \emph{not} contradicted by the anticorrelation between
continuum luminosity and equivalent width of quasar emission lines
known as the the Baldwin effect [\citealp{Baldwin1977}], as that
effect is strongest in high-ionization lines like \mbox{Ly\,$\alpha$}
and \mbox{C\,{\sc iv}}).  The constant Balmer-line width implies that
beaming does not change the optical continuum luminosity
significantly.  Hence, we expect FSRQs to have much higher values of
$R$ than other AGN purely due to relativistic beaming.  At the same
time, these objects may have also a high non-thermal fraction due to
beaming of the X-ray emission.  Hence, the trend for radio-loud
sources to be found at higher values of the non-thermal fraction may
be due to relativistic beaming affecting radio and X-ray emission, but
not the optical emission, in sources with the most powerful
relativistic jets (the trend of average $R$ increasing to the right in
Fig.~\ref{fiMapR}).  Unfortunately, we do not have spectral
information for our radio sources, so that we cannot remove the FSRQ
from our plot and check whether the remaining radio-loud objects still
show the same trend.

Nevertheless, it appears unlikely that the observed trends are
exclusively due to beaming and not connected to some intrinsic
differences.  First, the trend for the average $R$ to increase with
higher \emph{total} luminosity cannot be caused exclusively by
beaming, as there are high-$R$ objects whose total luminosity is more
than 50\% optical (thermal disc) emission.  Secondly, the radio-quiet
and radio-loud quasar population cannot be unified using orientation
and beaming alone (see \citealp{KellermannSramekSchmidt1989} and
references therein;
\citealp{RawlingsSaunders1991,MillerRawlingsSaunders1993}). In other
words, there is an intrinsic difference in addition to relativistic
beaming that gives some sources a much higher $R$ value than most
others. We verified that the observed gradient in average radio
loudness on the DFLD is also visible if we plot the fraction of radio
loud objects ($R>10$) on the DFLD.  Third, there is no correlation
between the radio and X-ray fluxes for the objects in our SDSS sample,
unlike in the samples considered by
\citet{GalbiatiCaccianigaMaccacaro2005} and
\citet{GrandiMalagutiFiocchi2006}.  Hence, even with a beamed
contribution to the X-rays, the fact that sources with high thermal
fraction are also more likely to have a high $R$ value also has to
arise from an intrinsic connection between thermal fraction and $R$
other than relativistic beaming.

In our analysis we use a fixed hydrogen column density for all
  quasars.  The used ROSAT fluxes are sensitive to changes in the
  column density. Thus, strong variations in the intrinsic column
  density of the quasars may change the position of the sources on the
  DFLD.  However, the majority of our sources is at moderate redshifts
  ($z \sim 1$) so the observed photons have been emitted as harder
  X-rays and are less sensitive to intrinsic obscuration. All our SDSS
  quasars have broad lines. According to the ``standard unification''
  of AGN, the intrinsic column density is small or negligible in these
  broad line sources (for absorption see
  \citealt{RichardsHallVandenBerk2003},  e.g., for the column density 
  \citealt{GranatoDaneseFranceschini1997}, e.g.). Nevertheless, we cannot
  rule out completely that absorption affects our DFLD. However, for
  this to become the dominant effect, there would have to be a strong
  anticorrelation between radio loudness and hydrogen column density,
  which we consider to be unlikely given that all objects in the
  sample are broad-line objects and selected on optical continuum
  magnitude.

Finally, neglecting the contribution of host galaxy starlight to the
SDSS optical quasar magnitudes has the tendency to let
lower-luminosity and lower-redshift sources appear too optically
luminous.  Furthermore, it artificially increases the disc fraction
and lowers the $R$ value. 
However, \citet{VandenBerkShenYip2006} have performed a spectral
decomposition of SDSS quasar spectra and find that only 20\% of the
objects considered have a host galaxy fraction of more than 50\%, and
hardly any have a host galaxy fraction of more than 90\%.  Thus, the
host galaxy contribution affects only a small fraction of objects, and
only in the lower left-hand part of the DFLD.  They will not affect
the high-luminosity sources that create most of the signal in the
upper right-hand part of the DFLD.  Nevertheless, it would clearly be
desirable to repeat our experiment including host galaxy corrections.

\subsection{Size-luminosity evolution of radio sources}
\label{s:disc.impact.size}

We only have radio data at a single observing frequency, 1.4\,GHz, and
with a rather large beam size of 5\arcsec, corresponding to physical
scales between 17\,kpc at $z=0.2$, our low-redshift cutoff, and
42\,kpc at $z=1.6$, where the angular diameter distance is maximal for
our chosen cosmology.  This implies that there will be some
contribution to our ``core'' luminosities from extended lobe
components in the FIRST beam.  This is not a problem for sources which
are large enough for the lobes to be clearly separated from the core
in FIRST imaging, since the SDSS catalog matching employs a rather
small matching radius (3\arcsec) for FIRST sources.  However,
\citet{KaiserDennettThorpeAlexander1997} have shown that lobes are
more luminous when the source is smallest, exacerbating the problem
for sources where the core cannot be discerned from the lobes based on
the FIRST imaging.  Nevertheless, these sources have lobes precisely
\emph{because} they have powerful jets, so that they must have a large
core radio flux, too.  Hence, any contribution from unresolved lobes
will have the effect of increasing the inferred value of $R$ only in
sources that have a high radio loudness anyway, but it will not
increase the radio loudness of sources without a jet.  In other words,
lobe contamination makes radio-loud sources even more radio-loud, but
it does not create false positives in regions of the DFLD where there
are no radio-loud sources.

\section{Discussion: Analogies with XRBs}
\label{s:disc}

We have shown in the previous section that quasars are more likely to
be radio-loud if the X-ray luminosity is large compared to the thermal
disc luminosity and/or the source is very luminous. This suggests the
radio loudness depends at least on 2 parameters: the total accretion
rate as well as a measure of the relative prominence of the
non-thermal X-ray component compared to the thermal disc
emission. This effect is similar to that observed in XRBs, which
motivated our development of the DFLD. Here we compare the behaviour
of AGN with XRBs.

\subsection{Monte-Carlo simulation of an XRB population DFLD}

For X-ray binaries, we can study the evolution of single objects in
the HID during their outbursts (e.g., Fig.~\ref{fiSketch}), while we
have to consider AGN as a population. Nevertheless, as DFLDs are
constructed as an analogue of the HIDs, the overall distribution of a
population of XRBs in an HID and the distribution of our AGN sample in
a DFLD should be similar. Indeed, XRBs and AGN populate the right side
of their diagrams at low luminosities and create a cloud of objects in
the soft and IM states further to the left at higher luminosities
\citep{FenderBelloniGallo2004}.  To compare XRBs and AGN in more
detail, we simulate a DFLD for an XRB population of 100 objects based
on our knowledge of the evolution of the disc and the power law
component in an outburst cycle of an XRB (a sufficient number of
observed HIDs is not readily accessible to us).  We fix the black-hole
mass of the simulated XRBs at $10 M_\odot$ \citep[see
e.g.,][]{Orosz2003}.

For each simulated outburst, we first choose the accretion rate at the
hard-to-soft state transition $\dot{M}_{\mathrm{crit}}$ randomly from
the range 5\%-100\% Eddington. We use $\Lrad/\Ld$ as measure of the
radio loudness to retain the same physical meaning as $\Lrad/\Lopt$
for AGN. During its outburst the source moves through its canonical
states, and we compute luminosities for each of the source components
using the prescription given in Appendix~\ref{app:XRBrecipe}.

\begin{figure}
\resizebox{8.7cm}{!}{\includegraphics{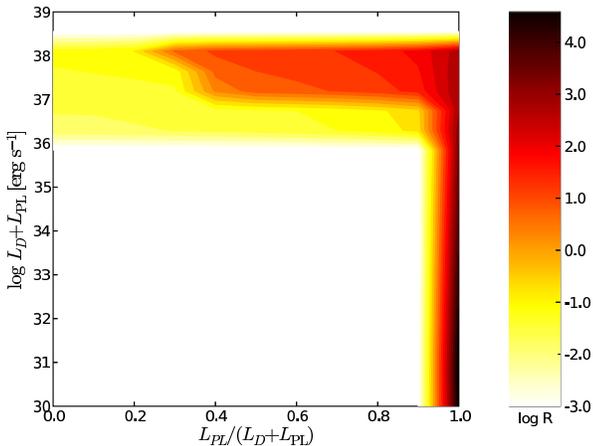}}
\caption{DFLD for a simulated sample of XRBs. We simulated 100 outbursts of an XRB.  }
\label{fiXRBs}
\end{figure}
Figure~\ref{fiXRBs} shows the simulated DFLD for 100 XRB
outbursts. The overall distribution of the simulated XRBs in the DFLD
is similar to the observed diagram for AGN. Also the radio loudness
distribution is similar for XRBs and AGN. The sources at the
right-hand edge of the diagram are the most radio-loud objects. In the
upper part, where the SDSS quasars are located in the AGN DFLD, we
find a gradient from high average $R$ in top right-hand corner to a
low average $R$ in the bottom left part, both for the simulated XRBs
and the observed AGN. For XRBs, the jet is quenched (we assumed
quenching by a factor 100) in the bottom left corner of the upper part
of the diagram (Fig.~\ref{fiMapR}); by contrast, all sources in the top right-hand corner
are extremely radio-loud. The gradient in radio loudness is smaller
for AGN, and we will discuss possible reasons below.  At low
luminosities, both XRBs and AGN are located in the right/power-law
dominated side of the diagram.

\subsection{Further evidence for common accretion states in XRBs
  and AGN} \label{se:EvidenceCommon}

\begin{figure}
\resizebox{8.7cm}{!}{\includegraphics{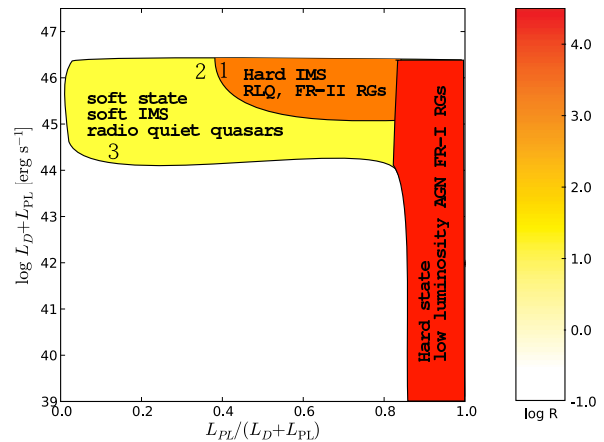}}
\caption{Sketch of the distribution AGN and XRB states (scaled to AGN masses) on the DFLD.
As discussed in the text, a lobe-dominated AGN could undergo a strong
change in the core properties, moving it from position ``1'' to
position ``2'', but the corresponding change in the lobe properties
would be delayed by at least the fading time of the lobe, so that it
would appear to be too radio-loud based on FIRST data.  By contrast,
AGN at position ``3'' should have no lobes.}
\label{fiSketch}
\end{figure}

The similarity of the DFLDs of XRBs and AGN futher supports the idea
that AGN have the same accretion states as XRBs. In both classes of
object, the hard state is characterized by the absence of a strong
disc component, a power-law component dominating the total luminosity,
and the presence of a jet. The jet power is likely to dominate over
the radiated power \citep{KoerdingFenderMigliari2006}. In the AGN zoo,
we classify LLAGN and FR~I radio galaxies as hard state objects,
\citep[as already suggested
by][]{Meier2001,FalckeKoerdingMarkoff2004,MaccaroneGalloFender2003,Ho2005,KoerdingFalckeCorbel2005}. The
hard IMS has some disc contribution, but the radio jet is still active
\citep{FenderBelloniGallo2004}. In the AGN population, we suggest that
this state corresponds to the radio-loud quasars. In the soft IMS and
the soft state in XRBs which are even more disc-dominated, the jet is
not observable in the radio. The corresponding AGN class are the radio
quiet quasars. In Fig.~\ref{fiSketch} we show a schematic diagram of
the distribution of the states in our DFLD.

For XRBs each cell of the DFLD can be associated with a given class,
as we have assumed that the BH masses of XRBs are similar and used a
deterministic evolution for all sources. Therefore, we find a fast
transition in our simulation from the radio-loud hard state and hard
IMS to the radio-quiet states (soft and soft IMS). In our AGN sample,
we have a large BH mass range and orientation will affect the observed
luminosities. Thus, the sharp transitions will be washed out. Even if
a cell was associated, for example, with the hard IMS, there would
still be a fraction of radio-quiet sources in this bin due to
different BH masses or source peculiarities.

With long radio observations, it is usually possible to find radio
emission from all radio-quiet quasars, and there are correlations
between radio and emission-line luminosity both for radio-quiet and
radio-loud objects
\citep[e.g.,][]{MillerRawlingsSaunders1993,FalckeMalkanBiermann1995,XuLivioBaum1999,HoPeng2001}.
The difference in radio loudness between radio-loud and quiet quasars
is typically around a factor 100
\citep[e.g.,][]{KellermannSramekSchmidt1989}.  For XRBs it is often
stated that the radio jet is quenched
\citep[e.g.,][]{FenderCorbelTzioumis1999} in the soft state. However,
the upper limits on the radio flux on high state objects are not yet
low enough to rule out radio emission from ``quenched'' XRBs at a
level equivalent to that in radio-quiet quasars (one of the best
studied cases is XTE~J1550$-$564 where \citet{CorbelKaaretJain2001}
find that the radio emission is quenched by at least a factor 50). 
The analogy between XRBs and AGN therefore suggests that
very deep radio observations of soft-state XRBs should detect radio emission.  

For XRBs we can measure how fast a sources moves through the DFLD. For
AGN it is not yet known how long a source spends in each accretion
state.  State changes in AGN might be triggered by the same process as
in XRBs, e.g., disc instabilities
\citep[cf.,][]{FrankKingRaine2002}. But already different XRBs or
different outbursts of the same source move through the diagram at
different speeds. Furthermore, environmental effects can also change
the external accretion rate dramatically (e.g, during a galaxy
merger), thus the timescales do not have to scale linearly with black
hole mass. These timescales will not play an important role for the
radio luminosity as a function of the position in the DFLD as long as
the radio emission responds to changes in the accretion rate on a
faster timescale than the timescale of a state change. Our radio data
is from the FIRST survey, i.e., high-angular resolution images at 1.4
GHz. Especially at higher redshifts the corresponding emitted
frequency is even higher and the radio emission is likely to originate
predominantly from the radio core. The emission region is therefore
fairly small (parsec-scale) and the radio flux will respond to any
state changes rather quickly ($\sim$ a few years) --- similar to XRB
case (here the response is even faster, $\sim$ hours).  However, the
large-scale radio structure (radio lobes) responds much more slowly.
It is therefore likely that some sources changing from the radio-loud
hard intermediate state to the radio-quiet soft state still have
strong radio lobes, even though the core is already in a radio-quiet
state. For a 100-kpc quasar jet, we see the core change up to 200 ky
{\it earlier} than the near lobe. The importance of this delay depends
on the sum of the fading time of the lobe and the typical lag between
the time when {\it we} see the core properties change, and the time
when the {\it lobe} receives the information that the core properties
have changed. On the DFLD in Fig.~\ref{fiSketch} we have marked three
positions with the numbers 1 to 3.  In the area marked as ``1'' we
expect to observe predominantly sources in the hard intermediate
state, i.e., sources with an active radio core and possibly strong
radio lobes. Some sources in the area ``2'' may have just changed
their accretion state to the soft state. Thus, some sources should
have a weak radio core but still have strong radio lobes. In
comparison, area ``3'' should be populated mainly with sources with no
lobes and only weak radio cores.

\subsection{Other DFLD trends: timing and jet Lorentz factor}

In XRBs, there are correlations between the source's position in the
DFLD and other observables than just the radio loudness.  Looking for
their presence in AGN will be a strong test of the AGN-XRB unification
scenario.

The timing features of XRBs depend on the accretion state of the
object and therefore on its position in the DFLD. Hard-state objects
on the right side of the diagram show strong flat-topped band limited
noise (see \citealt{Klis2004}). Soft-state objects on the lower left
part of the DFLD (area ``3'' in Fig.~\ref{fiSketch}) should mainly
show weak power-law noise. During the IMS strong quasi-periodic
oscillations (QPOs) are found. Near the transition from the hard to
the soft IMS, objects often show strong QPOs around 6 Hz \citep[type
B;][]{WijnandsHoman1999,BelloniHomanCasella2005}. With a linear mass
scaling, this would translate into a period of $\sim 200$\,d for
objects in area ``1'' and ``2''. However, as with the radio loudness
map, different black-hole masses and source peculiarities
(obscuration, beaming) will ``smear out'' the sources in the DFLD, so
there will not be a sharp transition as found in a single XRB
outburst.

Also the bulk Lorentz factor of XRB jets may depend on the position in
the DFLD \citep{FenderBelloniGallo2004}. It would be interesting to
see if this is also the case for AGN. To do so, one would need a large
sample of AGN with a measure of their bulk Lorentz factors. However,
reliable measurements of Lorentz factors are hard to obtain
\citep{Fender2003}. If we approach the jet line in the DFLD, the jet
in XRBs become instable and one observes rapid ejections
\citep{FenderBelloniGallo2004}. It is not yet clear whether this
behaviour is also found in AGN.  One caveat is that while hard-state
XRBs seem to have rather slow jets, it is thought that those in BL~Lac
objects are highly relativistic. According to the orientation
unification schema of AGN, the parent population of BL Lac objects are
low-luminosity FR~I radio galaxies, which are likely to be hard state
objects. Thus, it may be that the jet velocities are different for
stellar and supermassive black holes or that the observed slow jets
have a fast spine that has not yet been observed in XRBs.  A detailed
study is needed to decide whether the speeds of AGN jets are similar
to those of XRB jets.

\section{Conclusions}
\label{s:conc}

We have shown that the hardness-intensity diagrams used to diagnose
the states of XRBs can be generalised to disc-fraction/luminosity
diagrams (DFLDs), which can be used both for AGN and XRBs. The shape
of the distribution of low-luminosity AGN and SDSS quasars in this
diagram is similar to that of XRBs. Additionally, supermassive and
stellar-mass black holes show a similar dependence of radio loudness
on position in the DFLD. This suggests that radio loudness can only be
understood in context of a two-dimensional diagram, i.e., as function
of non-thermal fraction and luminosity, and is not directly related to
a single variable like the accretion rate or black-hole mass. The
similarity of the diagrams for AGN and XRBs supports the idea that AGN
have the same accretion states as XRBs.

At low luminosities, AGN and XRBs are always in their hard state,
which is characterized by strong X-ray and radio emission compared to
their disc emission. At higher luminosities, AGN can either be in
their radio-loud hard IMS (radio-loud quasars) or their soft state or
soft IMS (radio-quiet quasar). The position of an AGN in the DFLD
determines the probability with which it is in each of these states.
Thus, we find the same coupling between accretion flow/disc and jet
properties in AGN as in X-ray binaries.

To avoid all evolutionary and selection effects, it will be desirable
to repeat our investigation with a sample of AGN from a single narrow
redshift range, but with high dynamic range in luminosity and
well-determined masses.  Ideally, the sample would extend down to
LLAGN, which can to date only be found in the nearby universe, but not
in the same volume as large numbers of quasars.  Further strong tests
of the AGN-XRB unification and our proposed equivalence between XRB
states and members of the AGN zoo will be performed by looking for
correlations between DFLD position and other properties, such as QPO
frequencies.

\section*{Acknowledgements} 

The authors thank Tom Maccarone for helpful discussions. SJ was
supported by the Max-Planck-Institut f\"ur Astronomie through an Otto
Hahn Fellowship. We thank our referee for constructive comments.

Funding for the Sloan Digital Sky
Survey\footnote{\url{http://www.sdss.org}} (SDSS) has been provided by
the Alfred P. Sloan Foundation, the Participating Institutions, the
National Aeronautics and Space Administration, the National Science
Foundation, the U.S. Department of Energy, the Japanese
Monbukagakusho, and the Max Planck Society.  The SDSS is managed by
the Astrophysical Research Consortium (ARC) for the Participating
Institutions. The Participating Institutions are The University of
Chicago, Fermilab, the Institute for Advanced Study, the Japan
Participation Group, The Johns Hopkins University, the Korean
Scientist Group, Los Alamos National Laboratory, the
Max-Planck-Institute for Astronomy (MPIA), the Max-Planck-Institute
for Astrophysics (MPA), New Mexico State University, University of
Pittsburgh, University of Portsmouth, Princeton University, the United
States Naval Observatory, and the University of Washington.

\label{lastpage}

\bibliographystyle{mn2e}

\appendix

\section{SQL query for SDSS quasar sample}
\label{s:sql_app}

The query selects all objects from the SDSS spectroscopic database
that are classified with high confidence as a quasar, and, where
available, ROSAT and FIRST parameters.
\begin{verbatim}
select ...
from  specphoto as sp
left outer join rosat as ro on ro.objid = sp.objid
left outer join first as fi on sp.objid = fi.objid
where specclass in (3,4) 
and zconf >= 0.35 
and z between 0.2 and 2.5
\end{verbatim}

\section{Recipe for simulating an XRB outburst cycle}
\label{app:XRBrecipe}
Here, we describe our assumptions on the evolution of an XRB during a single outburst. A source is assumed to be in quiescence before the start of the simulation, i.e., we start the simulation with the rise of the outburst.
\begin{itemize}
\item Initial hard state: When a source starts its outburst, it is in
the hard state with its SED dominated by the power law
component. Therefore, we set $\Lpl =
\left(\frac{\dot{M}}{\dot{M}_{\mathrm{crit}}}\right)^2 0.1
\dot{M}_{\mathrm{crit}} c^2$, assuming a quadratic scaling of the
luminosity with accretion rate as expected for inefficient flow or jet
models (theoretical prediction: \citealt{NarayanYi1994}, empirical:
\citealt{KoerdingFenderMigliari2006}). The jet is active and the radio
luminosity $\Lrad \propto \dot{M}^{1.4}$
\citep{BlandfordKonigl1979,FalckeBiermann1995}. The disc is ususally
not directly visible in this state, we set its luminosity to 1 \% of
the PL component.
\item IMS: During the transition, the bolometric luminosity does not
change significantly \citep{ZhangCuiHarmon1997}. Thus, we set $\Lpl =
(1-\xi) \times 0.1\dot{M}_{\mathrm{crit}} c^2$ and $\Ld = \xi\times
0.1\dot{M}_{\mathrm{crit}} c^2$, where $\xi$ is the disc fraction
during the transition. The radio jet is active up to the jet line
\citep{FenderBelloniGallo2004}, which may vary with the critical
accretion rate $\dot{M}_{\mathrm{crit}}$. Higher-luminosity outbursts
have an active jet up to larger disc fractions $\xi$. We set the jet
line at $\xi = 0.57+ \frac{\dot{M}_{\mathrm{crit}} c^2}{10^{36}
  \mathrm{erg\,s}^{-1}}$,
which is normalised for the transition in GX~339-4 observed by
\citet{BelloniHomanCasella2005}. The jet power up to the jet line is
assumed to be constant and equal to the maximum jet power achieved in
the hard state (i.e., both the radiative and the kinetic luminosity
remain constant during the transition). We omit the relativistic
ejections which happen right at the transition to the radio-quiet
phase, as they are very short-lived. Their inclusion would yield a
higher radio flux right at the jet line. The strength and size on the DFLD
of this feature depends on the timescales of a source moving on the DFLD compared to
the timescale of a responds in the radio (see discussion in sec.~\ref{se:EvidenceCommon}).
\item In the soft state, the SED is dominated by the soft-spectrum,
efficiently radiating accretion disc: $\Ld = 0.1 \dot{M} c^2$. A hard
power law is often visible up to very high photon energies, we assume
$\Lpl = 0.01 \Ld$. The radio jet is quenched by a factor 100 compared to the hard state.
\item IMS: The transition back towards the hard state happens at lower
luminosity than the hard-to-soft transition. The hysteresis is smaller
for smaller transition luminosities; Cyg X-1, e.g., is nearly always
near the transition and does not show any hysteresis. Thus, we set
$\log (0.1 \dot{M}_{\mathrm{back}} c^2) = 37 - 0.5\,(\log
(0.1\dot{M}_{\mathrm{crit}} c^2)-37)$, where all luminosities are
measured in erg\,s$^{-1}$. For this transition we use a similar
prescription for the power-law luminosity as in the first transition:
$\Lpl = (1-\xi) \,0.1 \dot{M}_{\mathrm{back}} c^2$ and $\Ld = \xi\, 0.1
\dot{M}_{\mathrm{back}} c^2$, but the jet is assumed to be quenched
during the whole transition and is expected to restart only when the
source enters its hard state.
\item Declining hard state: same as initial hard state, just $\dot{M}_{\mathrm{crit}}$ is exchanged with $\dot{M}_{\mathrm{back}}$
\end{itemize}

\end{document}